\documentclass[aps,pra,floatfix,twocolumn,superscriptaddress]{revtex4}

\usepackage{graphicx}
\usepackage{graphics}
\usepackage{amssymb}
\usepackage{amsmath} 
\usepackage{epsfig}
\usepackage{latexsym}

\newcommand{\ket}[1]{\left|#1\right\rangle}
\newcommand{\bra}[1]{\langle #1|}
\newcommand{\braket}[2]{\left\langle #1|#2\right\rangle}
\newcommand{\op}[1]{\hat{#1}}
\newcommand{\eqrf}[1]{(\ref{#1})}

\begin{document}

\title{Entanglement of two-mode Bose-Einstein condensates}

\author{Andrew P. Hines}
\affiliation{Centre for Quantum Computer Technology,\\ Department of Physics, \\ The University of Queensland, St Lucia, QLD 4072, Australia}
\author{Ross H. McKenzie}
\affiliation{Centre for Quantum Computer Technology,\\ Department of Physics, \\ The University of Queensland, St Lucia, QLD 4072, Australia}
\author{Gerard J. Milburn}
\affiliation{Centre for Quantum Computer Technology,\\ Department of Physics, \\ The University of Queensland, St Lucia, QLD 4072, Australia}
\date{\today}

\begin{abstract}
 We investigate the entanglement characteristics of two general bimodal Bose-Einstein condensates - a pair of tunnel-coupled Bose-Einstein condensates and the atom-molecule Bose-Einstein condensate. We argue that the entanglement is only physically meaningful if the system is viewed as a bipartite system, where the subsystems are the two modes. The indistinguishibility of the particles in the condensate means that the atomic constituents are physically inaccessible and thus the degree of entanglement  between individual particles, unlike the entanglement between the modes, is not experimentally relevant so long as the particles remain in the condensed state. We calculate the entanglement between the two modes for the exact ground state of the two bimodal condensates and consider the dynamics of the entanglement in the tunnel-coupled case.
\end{abstract}

\maketitle

\section{Introduction}
In recent times, entanglement has come to be regarded as a physical resource which can be utilized 
to perform numerous tasks in quantum computation \cite{mixse}. This means that the creation and manipulation of entangled states is of significant interest in
quantum information and computation. On the other hand, the study of the entanglement characteristics of various condensed matter systems has been proposed to
provide new insights into quantum many-particle systems \cite{TON,maphd,cm1,cm2,cm3,cm4,shi}.

 One extensively studied condensed matter system \cite{leg,GMJ,bog,natmae} is that of a pair of 
tunnel-coupled Bose-Einstein condensates (BEC's). In the simplest model system, bosons are restricted to occupy one of two \emph{modes}, each of which is a BEC. A
dynamical scheme for engineering many-particle entanglement between the particles in such a system has been proposed by Micheli {\it et al} \cite{mpe2}. However,
since there is presently no definitive measure for entanglement between three or more subsystems, the amount of entanglement in the output state is not analyzed
quantitatively. Instead, this scheme aims to create states of a canonical form, whose entanglement content is based upon its inseparability.

 As argued in \cite{Benetal}, entanglement is only meaningful for multipartite systems whose 
Hilbert space can be viewed as a tensor product of two or more subspaces  corresponding to 
physical subsystems of the system. As always, what one regards as an
entangled state is, to some extent, a matter of how this decomposition of the system into 
subsystems is performed. One person's entangled state is not the same as
another's if they identify the subsystems differently. Entanglement can be 
said to be relative to the system decomposition \cite{zan}.

 In the case of a bimodal Bose-Einstein condensate investigated in \cite{natmae,mpe2}, the entangled subsystems were identified as the individual atoms in the condensate, and a mathematical measure of the multipartite entanglement proposed. By this measure certain states of the condensate were shown to be entangled. We argue here that the decomposition of the system in reference \cite{mpe2} into subsystems made up of individual bosons is not physically realizable, due to the indistinguishibility of the bosons within the condensates \cite{peres,li}. We argue that a more physically  relevant interpretation is to decompose the system into a bipartite system of the two modes. This idea is analogous to that of van Enk \cite{VE} concerning the entanglement of electromagnetic field modes, as opposed to the photons themselves. Since entanglement in bipartite systems is well understood, the entanglement between the modes of the system can be simply calculated. We demonstrate that while the states created in reference \cite{mpe2} are indeed entangled, the entanglement has a different character.

 As a further extension, we consider another two-mode system, the atom-molecule BEC  \cite{atm1,atm2,atm3,atm4,atm5,atm6,zol,don}. This systems has attracted significant interest since the entangled state is comprised of two chemically distinct components. We show that the entanglement between the atomic and molecular modes can be calculated analogously to the entanglement between the two modes of the tunnel-coupled BEC's.

\section{The Systems}\label{systems}
 The situation where a large number of interacting bosons are restricted to occupy the same two-dimensional single-particle Hilbert space is known as the {\it Josephson effect}. The Josephson effect can be described as either {\it external}, where the two single particle states, or modes, are separated spatially, or {\it internal} in which the two modes differ by some internal quantum number. Both the internal and external Josephson effects can be  described by the canonical Hamiltonian \cite{leg}
\begin{eqnarray}\label{ham}
\hat{H}_{J} & = & \frac{K}{8} \left(\op{N}_{A} - \op{N}_{B}\right)^{2} - \frac{\Delta\mu}{2} \left(\op{N}_{A} - \op{N}_{B}\right) \nonumber\\
  &   & - \frac{\mathcal{E}_{J}}{2}\left(\op{a}_{A}^{\dagger}\op{a}_{B} + \op{a}_{B}^{\dagger}\op{a}_{A}\right)
\end{eqnarray}
where $\op{a}_{A}^{\dagger},\op{a}_{B}^{\dagger}$ denote the single particle creation operators in the two modes ($A$ and $B$) respectively, and $\op{N}_{A} = \op{a}_{A}^{\dagger}\op{a}_{A}$, $\op{N}_{B} = \op{a}_{B}^{\dagger}\op{a}_{B}$ are the corresponding boson number operators. The parameter $\mathcal{E}_{J}$ is the single-atom tunneling amplitude, $\Delta \mu$ is the difference in the chemical potential between the wells and  $K$ corresponds to the atom-atom interaction. Here we only consider $K > 0$, corresponding to a repulsive interaction between atoms. The total particle number, $\op{N}_{A}+\op{N}_{B}$ is a conserved quantity and is set to the constant value $N$. If we add the constant term
\begin{equation}
\frac{K}{8} \left(\op{N}_{A} + \op{N}_{B}\right)^{2}
\end{equation}
the first term in the Hamiltonian \eqrf{ham} becomes
\begin{equation}
\frac{K}{4} \left(\op{N}_{A}^{2} + \op{N}_{B}^{2} \right)
\end{equation}
as we expect for repulsive $s$-wave scattering.

 This Hamiltonian (\ref{ham}) is in fact a two-site version of the Bose-Hubbard model which describes bosonic particles with repulsive interactions, hopping through a potential lattice \cite{fish,bh2}. In the Bose-Hubbard model, instead of two modes, there is an infinite lattice of potential wells (or modes) with coherent single-atom tunneling between nearest neighbor modes.

 A similar condensed matter system where there is the coupling of two BEC modes is that of atom-molecule Bose-Einstein condensate. In such a situation there exists coherent coupling between atomic and molecular BEC's respectively, which constitute the two modes of the system. The simplest Hamiltonian, recently studied by Vardi {\it et al}. \cite{vard}, which describes the atom-molecule BEC takes the form
\begin{equation}\label{hamatm}
\hat{H}_{AM} = \frac{\delta}{2} \op{a}^{\dagger}\op{a} + \frac{\Omega}{2}\left(\op{a}^{\dagger}\op{a}^{\dagger}\op{b} + \op{b}^{\dagger}\op{a}\op{a} \right)
\end{equation}
where $\op{a}^{\dagger}$ and $\op{b}^{\dagger}$ denote the creation operators for the atomic 
and molecular modes, respectively. $\Omega$ is a measure of the strength of the matrix elements for creation and destruction of molecules and $\delta$ is the
molecular binding energy in the absence of coupling. The total atom number $\op{N}_{atm} = \op{n}_{a} + 2\op{n}_{b}$, where $\op{n}_{a} = \op{a}^{\dagger}\op{a}$,
$\op{n}_{b} = \op{b}^{\dagger}\op{b}$, commutes with the Hamiltonian, so is a constant of the motion. Both Hamiltonians (\ref{ham}) and (\ref{hamatm}) have recently been shown by Zhou {\it et al}. \cite{zhoa,zho} to be exactly solvable in the context of the algebraic Bethe ansatz.

\section{Many-particle Entanglement}

For \emph{qubits} - two-dimensional systems represented by the states $\ket{0}$ and $\ket{1}$ - the 
canonical maximally entangled state is the Einstein-Podolsky-Rosen-Bohm (EPR) pair,
\begin{equation}
\frac{1}{\sqrt{2}}\left( \ket{00} + \ket{11} \right)
\end{equation}
also known as a Bell state, in reference to the inequalities established by Bell \cite{bell}. 
The tripartite analogue of this state is the Greenberger-Horne-Zeilinger-Mermin (GHZ) state
\begin{equation}
\frac{1}{\sqrt{2}}\left( \ket{000} + \ket{111} \right)
\end{equation}
while the corresponding $m$-partite state is given by
\begin{equation}
\frac{1}{\sqrt{2}}\left( \ket{0^{\otimes m}} + \ket{1^{\otimes m}} \right).
\end{equation}
These states are also known as $m$-particle Cat ($m$-Cat) states, in honour of Schr$\ddot{o}$dinger's cat. For systems with three or more subsystems, while most definitely entangled, we cannot say whether the Cat states are {\it maximally} entangled. Since there is no definitive measure for arbitrary multipartite entanglement, there is no clear notion of the structure of the maximally entangled sates in such systems.

The $d$-dimensional analog of the $2$-dimensional qubit is referred to as the \emph{qudit}. For qudits, represented by the set of states $\left\{ \ket{i} \right\}$ where $i = 0,\dots, d-1$, a Cat state would be
\begin{equation}\label{dcat}
\frac{1}{\sqrt{2}}\left(\ket{00} + \ket{(d-1)(d-1)} \right).
\end{equation}
By the standard measure of entanglement for bipartite systems (the {\it entropy of entanglement} which is discussed in section \ref{entmodes}) this is not the maximally entangled state. While state (\ref{dcat}) is entangled, a maximally entangled state is of the form
\begin{equation}\label{mqudit}
\frac{1}{\sqrt{d}}\sum_{i=0}^{d-1} \ket{ii}.
\end{equation}
While Cat states are the canonical maximally entangled states for systems consisting of two qubits, for higher dimensions and number of subsystems, the maximally entangled states correspond to uniform distributions over the tensor product basis.

In \cite{mpe2}, to determine the structure of the canonical entangled states, the system described by the 
Hamiltonian (\ref{ham}) was decomposed into $N$ subsystems consisting of the individual bosons, each with an internal degree of freedom described by a two dimensional Hilbert space spanned by the two states $\ket{A}$ and $\ket{B}$. In this description, the system is viewed as a collection  of $N$ single-qubit subsystems. These internal degrees of freedom can be used to define a two mode description just as the polarization degree of freedom of the electromagnetic field defines individual modes. In this case the annihilation and creation operators, appearing in equation (\ref{ham}), refer to single particle states distinguished by an internal degree of freedom rather than spatially localized single particle states discussed in this paper. However this does not change our point of view regarding the lack of physical significance of entanglement at the level of single atoms.
 In \cite{mpe2} it was argued that the maximally entangled state in this case is the $N$-Cat state, which is a coherent superposition state of all particles in mode $A$ and all particles in mode $B$,
i.e.
\begin{equation}\label{cats}
\frac{1}{\sqrt{2}}\left( \ket{A^{\otimes N}} + \ket{B^{\otimes N}}\right).
\end{equation}
 While it {\it cannot} be said that this is the maximally entangled state, it does indeed have some entanglement.

However there is a problem with this choice of subsystem partitioning.  By the nature of Bose-Einstein condensation, bosons within a condensate are {\it indistinguishable}. At no point can one make a physical measurement of the state of an individual particle in the condensate. For entanglement to exist between two systems the individual systems have to be distinguishable.  While it is easy to imagine quantum measurements sensitive to individual particles, such operations could not be realized in the laboratory \cite{peres}. While one can first remove individual particles from the condensate in order to measure them, the resulting state of the condensate is thereby changed, and it is unclear how the results of such measurements would reveal the multi-atom entangled state of the condensate prior to the removal of the measured particles. This implies that the decomposition into individual boson subsystems is not physically realizable and while one can still write the Hilbert space of the system as a tensor product of the Hilbert spaces of individual bosons, this is not an appropriate description for realizable measurements upon the condensate. In other words, the system of coupled BEC's is best viewed as a bipartite entangled system rather than as a collection of $N$ single-particle subsystems.
 
 Of course there is nothing to stop us from {\it calculating} the entanglement between indistinguishable particles according to some measure. However entanglement is a physical resource that enables useful tasks in quantum communication and computation. In all such tasks it is necessary that the entangled subsystems be distinguishable at some point in the protocols.  For systems described by the Josephson Hamiltonian (\ref{ham}) it cannot be said that there is {\it physically useful} entanglement between each individual boson when they exist in condensate. 

\section{Entanglement Between the Two Modes}\label{entmodes}

 Since the individual bosons are not physically accessible, distinguishable subsystems of the pair of tunnel-coupled 
BEC's described by (\ref{ham}), we need to consider other possible decompositions into subsystems if we are to investigate entanglement characteristics in this
system. While we cannot measure which mode of the coupled BEC's a specific particle is in, the occupation number of a given mode is a physical observable. The two modes, be they spatially separated, or differing in some internal quantum number, are clearly distinguishable subsystems. We can thus view the pair of coupled BEC's as a bipartite system of the two modes. It is relatively simple to investigate the entanglement between the modes since there is a unique measure of entanglement for two-component systems. Since the modes are distinguishable the entanglement between them is accessible and thus potentially useful for some quantum information or communication protocol. This had been demonstrated by Dunningham {\it et al.} \cite{dunn} who have proposed a scheme for entanglement swapping involving two pairs of tunnel-coupled BEC's. This is used to concentrate the entanglement between two modes.

 In this interpretation, while the entanglement involves many particles it is actually between the modes of the system. To illustrate this, consider the situation where we have just one particle in the system. In this scenario the modes can have occupation numbers of zero and one, so are spanned by the states $\ket{0}$ and $\ket{1}$. Consider the state
\[
\frac{1}{\sqrt{2}} \left(\ket{0}\ket{1} + \ket{1}\ket{0}\right).
\]
Clearly, with respect to the partition into modes, this single-particle state is entangled which 
implies we have entanglement with only a single particle. Analogous single-particle entanglement has been generated optically and used in a quantum teleportation protocol \cite{lee}.  

 The state of each mode is characterized by its occupation number. Because $N$ is constant, a general 
state of the system $\ket{\psi}$ can be written in term of the Fock states by
\begin{equation}\label{state}
\ket{\psi} = \sum_{n=0}^{N} c_{n}\ket{n}\ket{N-n}
\end{equation}
where $c_{n}$ are complex coefficients, i.e., $n$ bosons in mode $A$ implies there are $N-n$ bosons in mode $B$.

The standard measure of entanglement of pure states of bipartite systems is the {\it entropy of entanglement}, which is the von Neumann entropy of the reduced density operator of either of the subsystems \cite{pres,man}. The reduced density operator of a subsystem is found by {\it tracing out} the other subsystem via the {\it partial trace}. If $\rho$ is the density operator describing some state of a bipartite system, the reduced
density operator for subsystem $A$ is defined by
\begin{equation}  
\rho_{A} = \text{Tr}_{B} \left(\rho\right)
\end{equation}
where $\text{Tr}_{B}$ is the partial trace over subsystem B. The entropy of entanglement is then given by
\begin{eqnarray}\label{ent}
E(\rho_{A}) & = & -\text{Tr}\left(\rho_{A} \log \left(\rho_{A}\right)\right) \\
 & = & -\sum_{k} \lambda_{k} \log\left(\lambda_{k}\right)
\end{eqnarray}
where the logarithm is taken in base 2, and $\left\{ \lambda_{k}\right\}$ are the set of eigenvalues of the 
reduced density operator, $\rho_{A}$. The value of $E$ varies between $0$, for separable product states, to a maximum of $\log d$ (where $d$ is the dimension of
the Hilbert space of the subsystem) for maximally entangled states corresponding to a completely mixed density operator.

 The entropy of entanglement can be calculated from the reduced density operator of either of the subsystems without loss of generality. This follows from the {\it Schmidt decomposition} of pure states, which demonstrates that the eigenvalues of the reduced density operators of the two subsystems are identical ( pg. 109 of \cite{man}).

{\bf Schmidt decomposition}: For any pure state $\ket{\psi}$ of a bipartite composite system there exist 
orthonormal states $\ket{i_{A}}$ for subsystem $A$ and orthonormal states $\ket{i_{B}}$ for subsystem $B$ such that
\begin{equation}
\ket{\psi} = \sum_{i} \lambda_{i}\ket{i_{A}}\ket{i_{B}},
\end{equation}
 where $\lambda_{i}$ are non-negative, real numbers known as Schmidt coefficients, satisfying $\sum_{i} \lambda_{i}^{2}=1$. It is easy to see from the Schmidt decomposition that the reduced density operators for the two subsystems are, respectively, $\rho_{A} = \sum_{i}\lambda_{i}^{2}\ket{i_{A}}\bra{i_{A}}$ and $\rho_{B} = \sum_{i} \lambda_{i}^{2}\ket{i_{B}}\bra{i_{B}}$ which have identical eigenvalues.

 Using the Fock basis, from equation (\ref{state}) the density operator describing a general state of the system is given by
\begin{eqnarray}\label{rho}
\rho = \ket{\psi}\bra{\psi} = \sum_{m,n = 0}^{N} c_{m}c_{n}^{*} \ket{m}\ket{N-m}\bra{n}\bra{N-n}.
\end{eqnarray}

Taking the partial trace with respect to mode B yields the reduced density operator for mode A,
\begin{eqnarray}\label{redop}
\rho_{A}& = & \text{tr}_{B} \left(\rho\right) \nonumber\\
	& = & \sum_{m,n,k = 0}^{N} c_{m}c_{n}^{*} \ket{m}\bra{n}\braket{k}{N-m}\braket{N-n}{k} \nonumber \\
	& = & \sum_{n=0}^{N} |c_{n}|^{2} \ket{n}\bra{n}.
\end{eqnarray}

 From expression (\ref{redop}) we can see that the reduced density operator in this case is diagonal in the Fock basis and the eigenvalues are simply $\lambda_{i} = |c_{i}|^{2}$. Thus the entropy of entanglement 
between the two modes of the coupled BEC's is given
by
\begin{eqnarray}\label{enteig}
E\left(\rho_{1}\right) & = & - \sum_{n=0}^{N}|c_{n}|^{2} \log |c_{n}|^{2}.
\end{eqnarray}

 To determine the maximally entangled state, expression (\ref{enteig}) can be optimized with respect to $x_{n} = |c_{n}|^{2}$ by imposing the normalization condition $\sum_{n=0}^{N}  |c_{n}|^{2}=1$ with a Lagrange multiplier, $\mu$ i.e. we maximize
\begin{equation}\label{lang}
L = -\sum_{n=0}^{N} \left(x_{n}\log x_{n} - \mu x_{n}\right) +\mu.
\end{equation}
Differentiating with respect to $x_{n}$ gives
\begin{equation}
\frac{\partial L}{\partial x_{n}} = \mu - \log x_{n} - \frac{1}{\ln 2} = 0
\end{equation}
which implies
\begin{equation}
x_{n} = 2^{\mu -\frac{1}{\ln 2}}
\end{equation}
for all $n$. From the normalization condition 
\begin{eqnarray*}
x_{n} =& \frac{1}{N+1}, &\forall n \\
\Rightarrow |c_{n}|= &\frac{1}{\sqrt{N+1}}, & \forall n.
\end{eqnarray*}
So a state with maximum entanglement will have coefficients
\begin{equation}
c_{n} = \frac{\text{e}^{i\theta_{n}}}{\sqrt{N+1}}
\end{equation}
where $\theta_{n}$ is some phase angle. This corresponds to a completely mixed density operator, 
as expected for a state with maximal entanglement. Thus we can express the canonical maximally entangled state, $\ket{MES}$ for the system described by the
Josephson Hamiltonian (\ref{ham}) as
\begin{equation}\label{maxent}
\ket{MES} = \frac{1}{\sqrt{N+1}}\sum_{n=0}^{N} \ket{n}\ket{N-n}.
\end{equation}

 From equation (\ref{enteig}), the maximal entanglement is
\begin{eqnarray}\label{maxenta}
E_{max} & = & -\sum_{n=0}^{N} \frac{1}{N+1}\log\left(\frac{1}{N+1}\right) \nonumber\\
	& = & - \log\left(\frac{1}{N+1}\right) \nonumber\\
	& = & \log\left(N+1\right)
\end{eqnarray}
As mentioned previously, this is what is expected for the maximum entanglement, since the dimension of the 
Hilbert space of the individual modes is $N+1$ (see page 510 of \cite{man}).\\

\subsection{Entanglement of the Ground State}\label{gsent}
 As mentioned in Section \ref{systems}, the systems consisting of a pair of tunnel-coupled BEC's is the simplest system described by the Bose-Hubbard model and corresponds to a lattice potential with just two sites. For the Bose-Hubbard model, in the limit of an infinite lattice, there is a quantum phase transition where the ground state changes from superfluid phase to the Mott insulator phase \cite{fish}. Such a transition from the Mott insulator to superfluid phase was recently experimentally observed by Greiner {\it et al}. \cite{fmot}. 

In the Mott insulator state, particles tend to be localized at the individual lattice sites with no phase coherence across the lattice, whereas in the the superfluid state, each atom is spread over the entire lattice and there exists long-range phase coherence across the lattice. This transition from the Mott insulator to the superfluid state occurs as the ratio of the coupling between lattice sites to the interaction strength increases. As long-range coherences in quantum systems are intrinsically linked to entanglement, it is of interest to quantify the entanglement in the system in relation to this transition. Since there is no measure for the entanglement in systems consisting of three or more subsystems, the two mode system here is the only Bose-Hubbard model for which we can currently give a complete description of the entanglement.

 Making the two modes identical (by setting the bias, $\Delta\mu$ to $0$), the Hamiltonian (\ref{ham}) was diagonalised numerically for increasing values coupling to interaction ratio, $\frac{\mathcal{E}_{J}}{K}$, and the entanglement of the ground state was determined via equation (\ref{redop}).
\begin{figure}[ht]
\begin{center}
\scalebox{0.45}{\includegraphics{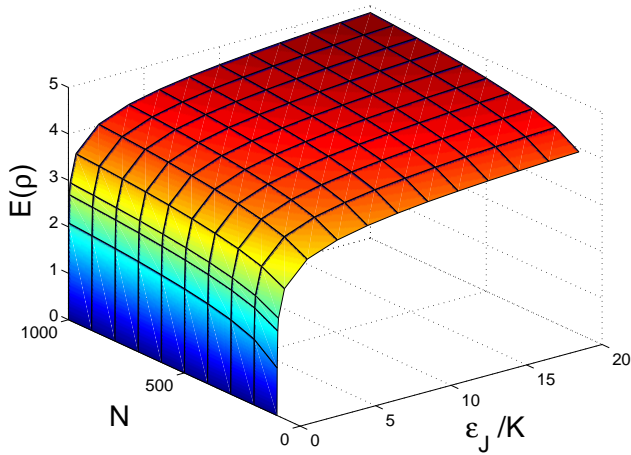}}
\end{center}
\caption{The variation of the entropy of entanglement of the ground state for differing particle number, $N$, and increasing coupling to atom-atom interaction ratio, $\frac{\mathcal{E}_{J}}{K}$.}
\label{3d}
\end{figure}

Figure \ref{3d} shows the results of this analysis for different values of the total particle number, $N$.
 Since $\mathcal{E}_{J}$ is the tunneling parameter, the larger its value the stronger the interaction between the modes of the system. As such, it is intuitive that that for no coupling, the entanglement between the modes is zero and as the coupling increases, the entanglement between the modes in the ground state increases.
\begin{figure}[ht]
\begin{center}
\scalebox{0.45}{\includegraphics{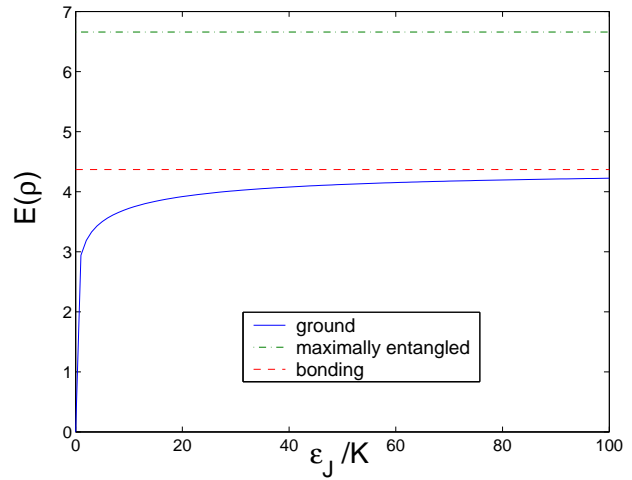}}
\end{center}
\caption{The variation of the entropy of entanglement of the ground state for increasing coupling to atom-atom interaction ratio, $\frac{\mathcal{E}_{J}}{K}$for $N=100$.}
\label{entgs}
\end{figure}

 The entanglement asymptotically approaches a constant value as the ratio  
$\frac{\mathcal{E}_{J}}{K} \rightarrow \infty$, which is illustrated more clearly in figure \ref{entgs}, which shows the results for $N=100$. Now we consider the two extreme parameter values. Firstly for $\mathcal{E}_{J} = 0$, the ground state will have an equal, fixed number atoms in each of the two modes and is therefore the localized state,
\begin{equation}\label{mott}
\ket{\psi_{loc}} = \ket{\frac{N}{2}}\ket{\frac{N}{2}}.  
\end{equation}  
Clearly this state has zero entanglement. For $K=0$, the Hamiltonian consists of just the tunneling term
\begin{equation}\label{Htun}
H_{tun} = - \frac{\mathcal{E}_{J}}{2}\left(\op{a}_{A}^{\dagger}\op{a}_{B} + \op{a}_{B}^{\dagger}\op{a}_{B}\right)
\end{equation}
and it is easy to show that the ground state of such a Hamiltonian for a single particle is the \emph{bonding} state
\begin{equation}
 \ket{+}=\frac{1}{\sqrt{2}}\left(a_{A}^{\dagger}+a_{B}^{\dagger}  \right)\ket{0}\ket{0}
\end{equation}
where $\ket{0}\ket{0}$ is the vacuum state. So in the ground state for $K=0$ each individual particle is in the bonding state and the state of the system is the $N$-particle analogue of the bonding state
\begin{equation}\label{sf}
\ket{\psi_{+}} =  \frac{1}{\sqrt{2^{N}N!}}\left(a_{A}^{\dagger}+a_{B}^{\dagger}  \right)^{N}\ket{0}\ket{0}.
\end{equation}
This state is the two-site analogue of the superfluid phase, with each atom being spread over the two modes. From equation (\ref{enteig}), the corresponding entanglement for this state is
\begin{equation}\label{bondent}
E(\rho_{+}) = -\sum_{n=0}^{N}\frac{1}{2^{N}}{N \choose n}\log\left(\frac{1}{2^{N}}{N \choose n} \right).
\end{equation}

Thus for zero coupling, the ground state is the localized state \eqrf{mott} which has no entanglement. As soon as the coupling begins to increase, the entanglement between the modes increases rapidly, as the occupation number of the modes is no longer exact and fluctuations in the phase decrease. As the tunneling amplitude continues to increase, the entanglement asymptotically approaches a maximum value, given by expression (\ref{bondent}), which corresponds to the bonding state. In this state, the occupation number fluctuations are large, resulting in phase coherence between the modes which is characterized by the high entanglement.

\begin{figure}[ht]
\begin{center}
\scalebox{0.45}{\includegraphics{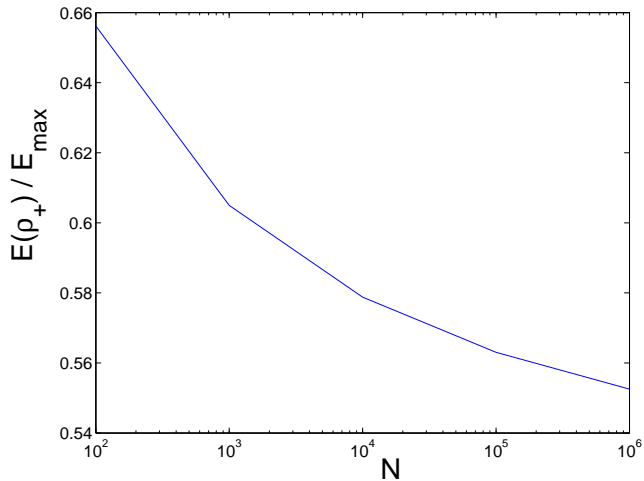}}
\end{center}
\caption{The ratio of the entanglement of the bonding state to the maximal entanglement, $E(\rho_{+}) / E_{max}$ for increasing particle number, $N$.}
\label{ratio}
\end{figure}

 Figure \ref{ratio} shows that the maximum entanglement in the ground state is much less than the maximal entanglement of the system, $\log(N+1)$.

However it appears the ratio of the entanglement of the two states remains finite as $N\rightarrow \infty$.\\

\subsection{Comparison of Entanglement}
The dynamical schemes developed in Ref.\cite{mpe2} aim to create states of the form (in the Fock basis)
\begin{eqnarray}\label{catd}
\ket{\mathrm{Cat}(D)} &= & \frac{1}{\sqrt{2}} \left( \left|\frac{N+D}{2}\right\rangle\left|\frac{N-D}{2} \right\rangle + \right. \nonumber\\
& & \left. \left|\frac{N-D}{2} \right\rangle\left|\frac{N+D}{2} \right\rangle \right)
\end{eqnarray} 
where $D = N_{A}-N_{B}$. It was argued that the amount of entanglement in the state (\ref{catd}) is characterized by the ``distance'', $D$. For $D=0$ the state is separable and thus not entangled, while for $D=N$, $\ket{\mathrm{Cat}(N)}$ is equivalent to (\ref{cats}) so is maximally entangled (which in this decomposition is still an incorrect assumption). However since motivation for the definition of these states was the decomposition of the system into individual particle subsystems, these states will have different entanglement characteristics when analyzed with the BEC modes as the subsystems. From equation (\ref{enteig}) it is clear that for $D=0$ the state (\ref{catd}) is separable and thus unentangled. However, for $D > 1$, we have  $E(\ket{\mathrm{Cat}(D)}\bra{\mathrm{Cat}(D)}) = 1$ - the amount of entanglement between the modes is the same independent of the value for $D$ ($>1$). The initial state prepared for the dynamical scheme of \cite{mpe2} is the $N$-particle bonding state (\ref{sf}). According to expression (\ref{bondent}), the entanglement between the modes in this state is actually greater than the entanglement in the final ideal output state, given by expression (\ref{catd}) with $D=N$. From this observation, it seems that the dynamical process outlined in \cite{mpe2} \emph{destroy} entanglement between the modes. To gain a better understanding of this dynamical process with respect to the modal decomposition, we consider how the entanglement between the modes varies during the evolution.

\subsection{Dynamics of Entanglement}\label{dyn}

In studying the dynamics of the system of tunnel-coupled BEC's, we express the the Hamiltonian (\ref{ham}) in the pseudo-angular momentum representation, introduced in \cite{GMJ}, used in \cite{mpe2}. In this representation we define the three angular momentum operators
\begin{eqnarray}
\op{J}_{z} & = & \frac{1}{2}\left( \op{N}_{B} - \op{N}_{A}\right) \label{jz}\\
\op{J}_{x} & = & \frac{1}{2}\left( \op{a}_{A}^{\dagger}\op{a}_{B} + \op{a}_{B}^{\dagger}\op{a}_{A} \right) \label{jx}\\
\op{J}_{y} & = & \frac{i}{2}\left( \op{a}_{A}^{\dagger}\op{a}_{B} - \op{a}_{B}^{\dagger}\op{a}_{A} \right) \label{jy}
\end{eqnarray}

which have the canonical commutation relations $[\op{J}_{x},\op{J}_{y}] = i \op{J}_{z}$ (and cyclic permutations). The Casimir invariant is easily seen to be

\begin{equation} 
\op{J}^{2} = \frac{N}{2}\left(\frac{N}{2} + 1\right).
\end{equation}
In this way the tunnel-coupled pair of BEC's system is analogous to an angular momentum model with total angular momentum $j = \frac{N}{2}$. The Hamiltonian (\ref{ham}) (with $\Delta\mu = 0$) can thus be rewritten as
\begin{equation}\label{hamang}
\op{H}_{J2} = \chi \op{J}_{z}^{2} - \Omega\op{J}_{x}
\end{equation}

where we neglect constant energy shifts, $\chi = \frac{K}{2}$ and $\Omega = \mathcal{\epsilon}_{J}$. This Hamiltonian \eqrf{hamang} is slightly different from that defined in \cite{mpe2} (and \cite{GMJ}). In these references, the $\op{J}_{x}$ term, which corresponds to the tunneling term, was added rather than subtracted (we assume only positive parameter values). However, this does not change the eigenstates of the system, but does reverse their order in terms of energies i.e. the ground state in the addition case is the highest excited state in the subtraction case. This means that the results for the dynamics from \cite{mpe2} can still be applied to Hamiltonian \eqrf{hamang} by using a different initial state, as we will discuss below.

 In the angular momentum representation, states can be expanded in terms of the $\op{J}_{z}$ eigenstates, $\ket{j,m}_{z}$, where $-j \leq m\leq j$. In this basis there is no indication of the underlying subsystem structure of the system. It is interesting to note that in terms of the Fock basis we have 
\[
\ket{j,m}_{z} \equiv \ket{N - 2m}\ket{N+2m}.
\]
This implies that for any state
\begin{eqnarray}
\ket{\psi} & = & \sum_{n=0}^{N} c_{n}\ket{n}\ket{N-n} \\
					 & = & \sum_{n=0}^{N} c_{n}\ket{j,n-N/2}_{z}
\end{eqnarray}
meaning that the entanglement between the modes can be calculated from the coefficients in the angular momentum basis.

 In reference \cite{mpe2}, a semi-classical model of Hamiltonian \eqrf{hamang} was used to determine the optimal parameter values and time scale to create states of the form of expression (\ref{catd}) with $D=N$ ($\ket{\mathrm{Cat}(N)}$), from the evolution of a given initial state. In the angular momentum representation 
\begin{equation}\label{catang}
\ket{\mathrm{Cat}(N)} = \frac{1}{\sqrt{2}} \left( \ket{j,-j}_{z} + \ket{j,j}_{z} \right)
\end{equation}

 From this analysis it was argued that using the critical parameter ratio
\begin{equation}\label{parat}
\frac{2\Omega}{\chi N} = 1
\end{equation}
in the evolution of the initial state over time, $t_{c} = \ln(8N) / \chi N$ (where time is in units of $\hbar$), could create states of the form (\ref{catang}).

 The initial state given in \cite{mpe2} was the maximal weight state of $\op{J}_{x}$. This corresponds to the bonding state, \eqrf{sf}. For the Hamiltonian above, the similar results will be achieved using the {\it minimal} $\op{J}_{x}$ weight state which is

\begin{equation}
\ket{\psi(0)} = \ket{j,-j}_{x} \equiv \frac{1}{\sqrt{2^{N}N!}}\left(a_{A}^{\dagger}-a_{B}^{\dagger}  \right)^{N}\ket{0}\ket{0}.
\end{equation}  

In other words

\begin{equation}
\ket{\psi(t_{c})} = e^{-i \op{H}_{J2} t_{c}}\ket{\psi(0)} \approx \ket{\mathrm{Cat}(N)}.
\end{equation}

 Using this parameter ratio and the initial state, the dynamics were investigated by numerically integrating the Schr$\ddot{\mathrm{o}}$dinger equation in the eigenbasis of $J_{z}$. 
\begin{figure}[ht]
\begin{center}
\scalebox{0.45}{\includegraphics{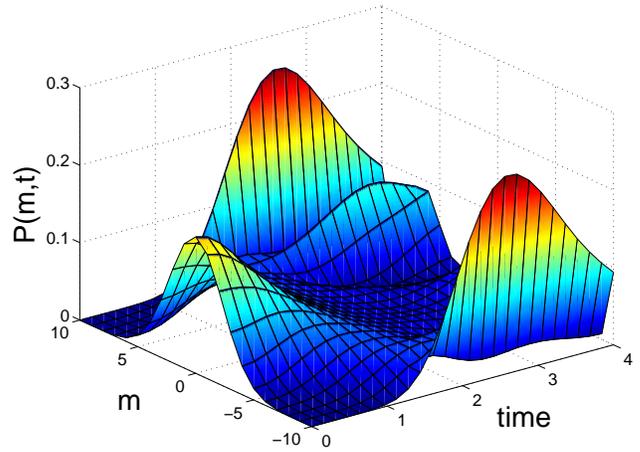}}
\end{center}
\caption{The evolution of the $J_{z}$ distribution for $j = 10$ ($N = 20$) with $2\Omega / \chi N  =1$. Note that at time $t_{c} \approx 2.5376$ the distribution is peaked at the two extreme $m$ values, corresponding to an approximate {\it Cat} state as shown in \cite{mpe2}. Again, the time is in units of $\hbar$.} 
\label{dist}
\end{figure}

 Figure \ref{dist} shows a plot of the evolution of the exact $J_{z}$ distribution, $P(m,t) = |_{z}\langle{j,m}\ket{\psi(t)}|^{2}$. This concurs with the results of \cite{mpe2}, showing that at time $t_{c}$ the final state is approximately of the form of the Cat state given in expression (\ref{catang}).

\begin{figure}[ht]
\begin{center}
\scalebox{0.45}{\includegraphics{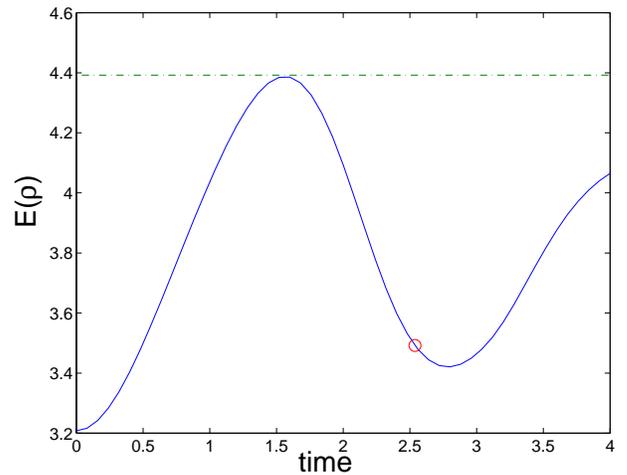}}
\end{center}
\caption{The evolution of the entanglement. The dashed line shows the maximal entanglement, while the circle indicates the entanglement at time $t_{c}$ (time in units of $\hbar$.)} 
\label{event}
\end{figure}

 The corresponding entanglement between the modes over the evolution is shown in figure \ref{event}. The state at time $t_{c}$ is not an exact {\it Cat} state, and as such the entanglement at $t_{c}$ was found to be greater than the initial entanglement. This differs from our earlier observation that the entanglement in the initial state is destroyed. This would be true if the state at time $t_{c}$ was exactly $\ket{\mathrm{Cat}(N)}$ (\ref{catang}).

 However, from figure \ref{event} it is easy to see that the maximum entanglement is not reached at this critical time but somewhat earlier, and at $t_{c}$ the entanglement between the modes is in the region of a local minima. As we can see, in the early stages of the evolution, the entanglement between the modes approaches the maximal entanglement of the system. Relating this to figure \ref{dist} it is easy to see that the maximum entanglement is approached as the original peak in the probability distribution flattens over the evolution, approaching a completely even distribution, corresponding to the maximally entangled state (\ref{maxent}). As the evolution continues, the distribution begins to peak at the extremes and the entanglement decreases. 

 In terms of modal entanglement, the dynamical scheme proposed in \cite{mpe2} can still be used to create a close to maximally entangled state, over a shorter time period than for the creation of the inaccessible many-particle entangled state.

 It is of interest to note that the critical parameter ratio (\ref{parat}) is the same as that found by Milburn {\it et al}. \cite{GMJ} in regards to a transition in the dynamics from self-trapping to delocalization. For an initial condensate localized in one mode, when $\frac{2\Omega}{\chi N} > 1$, the condensate distribution  will remain localized within the mode as it evolves. For $\frac{2\Omega}{\chi N} < 1$, the evolution results in a delocalization of the condensate distribution between the two modes.

\section{The Atom-Molecule BEC}

 The atom-molecule BEC described by Hamiltonian (\ref{hamatm}) is a similar system to that of the tunnel-coupled pair of BEC's. In neither system can we consider the individual particles (the individual atoms and molecules) as separate, distinguishable subsystems but both consist of the coherent coupling of two distinct BEC's. In the atom-molecule BEC, the two modes of the system do not differ spatially or by some internal quantum number but are rather two chemically distinct components. Nonetheless, the determination of the entanglement between the atomic and molecular
modes is analogous to the calculations above for the tunnel-coupled BEC's.

 As before, the state of each mode is characterized by it's occupation number, however in the case of the atom-molecule BEC the set of Fock states spanning the Hilbert space of the system depends upon whether the total number of atoms, $N_{atm}$, is even or odd. In
the case of an even $N_{atm}$, a general state, $\ket{\chi}$ of the system can be expanded as
\begin{equation}\label{atmse}
\ket{\chi} = \sum_{n=0}^{M} d_{n}\ket{2n}\ket{M-n}
\end{equation}
 where $M = N_{atm} / 2$, while for $N_{atm}$ odd, the general state $\ket{\phi}$ can be expressed as
\begin{equation}\label{atmso}
\ket{\phi} = \sum_{n=0}^{M} d_{n}\ket{2n+1}\ket{M-n}
\end{equation}
where in this case, $M = \left(N_{atm}-1\right) / 2$ and the $\left\{d_{n}\right\}$ are complex coefficients defining the state. In analogy with expression (\ref{redop}) for the reduced density operator for the tunnel-coupled BEC's, the reduced density operator for a general state of the atom-molecule BEC is given by
\begin{equation}\label{redopatm}
\rho_{b} = \sum_{n=0}^{M} |d_{n}|^{2}\ket{M-n}\bra{M-n}
\end{equation}
where the partial trace has been taken with respect to the atomic mode and $M$ is defined as above for even and odd total atom number, $N_{atm}$. Thus the entropy of entanglement between the atomic and molecular modes is given by equation (\ref{enteig}), the same expression as for the tunnel-coupled BEC's, where $N$, the total particle number, is replaced by $M$ as defined above i.e.
\begin{eqnarray}\label{enteigatm}
E\left(\rho_{b}\right) & = & - \sum_{n=0}^{M}|d_{n}|^{2} \log |d_{n}|^{2}.
\end{eqnarray}
Since the dimension of the subspace of the modes is $M$, the maximally entangled states, analogous to (\ref{maxent}), are
\begin{equation}\label{maxente}
\ket{MES_{even}} = \frac{1}{\sqrt{M+1}}\sum_{n=0}^{M} \ket{2n}\ket{M-n},
\end{equation}
for $N_{atm}$ even, and
\begin{equation}\label{maxento}
\ket{MES_{odd}} = \frac{1}{\sqrt{M+1}}\sum_{n=0}^{M} \ket{2n+1}\ket{M-n},
\end{equation}
for $N_{atm}$ odd and will have entanglement $\log\left(M+1\right)$.

\begin{figure}[ht]
\begin{center}
\scalebox{0.45}{\includegraphics{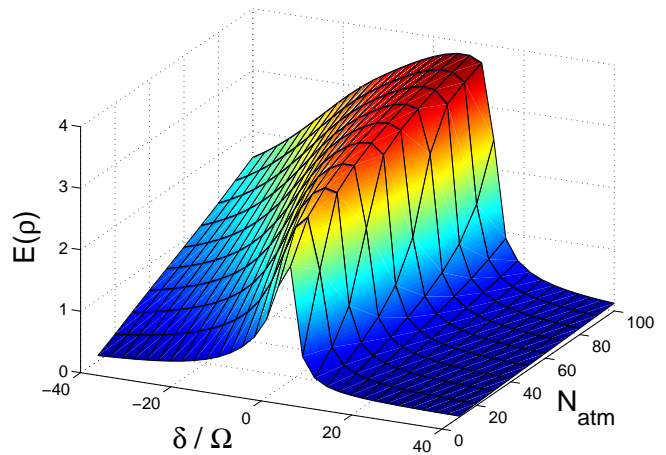}}
\end{center}
\caption{The entropy of entanglement of the ground state of the atom-molecule BEC for increasing values of the ratio $\frac{\delta}{\Omega}$ and atom number, $N_{atm}$.} 
\label{atment}
\end{figure}

\begin{figure}[ht]
\begin{center}
\scalebox{0.45}{\includegraphics{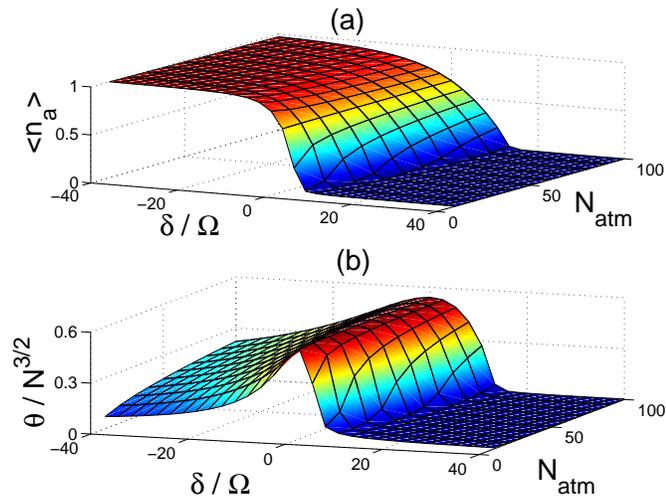}}
\end{center}
\caption{The average atomic occupation number (a) and the coherence correlator (b) for the ground state of the atom-molecule BEC. Note that both the average atomic occupation number and the coherence correlator have been scaled using the total atom number, $N$.} 
\label{atmboth}
\end{figure}

 Following the same numerical analysis as in section, \ref{gsent} figure \ref{atment} shows the results for 
the variation in the entanglement of the ground state of the atom-molecule BEC for differing values of the ratio of the parameters, $\frac{\delta}{\Omega}$ and total number of atoms, $N_{atm}$.
 To relate the entanglement structure of the ground state shown in figure \ref{atment} to some physical 
properties of the system we need to consider  other properties of the ground state for increasing total atom number and parameter ratio $\frac{\delta}{\Omega}$.
Zhou {\it et al}. \cite{zho} considered the two zero temperature correlations $\langle \op{n}_{a} \rangle$, the average atomic occupation number and the {\it
coherence correlator}, $\theta = \left\langle \op{a}^{\dagger}\op{a}^{\dagger}\op{b} + \op{b}^{\dagger}\op{a}\op{a} \right\rangle$.
 Figure \ref{atmboth} shows the results for the average atomic occupation number and the 
coherence correlator for the same parameter ranges used for the entanglement calculations. 

We should note that the results here using direct numerical diagonalization of the Hamiltonian 
concur with the results found by Zhou {\it et al}. \cite{zho} utilizing the exact solution.

 From figure \ref{atmboth}a it can be seen that the maximum entanglement in the ground state occurs 
where the average atomic occupation is comparative to the average molecular occupation. 
As indicated in \cite{zho}, the threshold coupling for the formation of a
predominantly molecular BEC is $\frac{\delta}{\Omega} \approx 1.4\sqrt{N_{atm}}$. 
In the limit of large $N_{atm}$, the threshold for the molecular BEC is in fact a quantum phase transition. 
\begin{figure}[ht]
\begin{center}
\scalebox{0.45}{\includegraphics{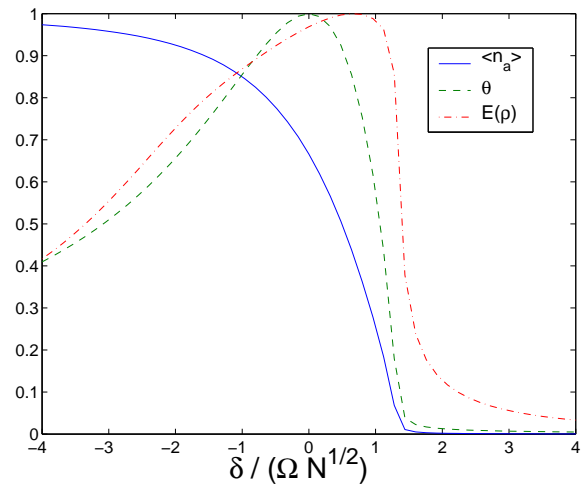}}
\end{center}
\caption{The average atomic occupation number, coherence correlator and the entanglement for the ground state of the atom-molecule BEC, for $N_{atm}=100$. All three properties have been scaled with respect to their maximum value so as to compare the characteristics of these properties.} 
\label{scaled}
\end{figure}
Figure \ref{scaled} shows the comparative results for the average atomic occupation number, the coherence correlator and the entanglement for $N_{atm}=100$. That the entanglement is \emph{not} maximal at the quantum critical point is quite different to the behavior of the transverse Ising model, studied in references \cite{TON,ost}. The entanglement characteristics of the transverse Ising model are, of course, much more complicated, since it consists of many distinguishable subsystems. In \cite{TON} it was conjectured that, in the sense of \emph{entanglement sharing} - how much two-party entanglement can be distributed amongst a given number of parties - the ground state was maximally entangled at the critical point. At the critical point the ground state saturates the bounds of entanglement sharing. While the ground state entanglement is not maximal at the critical point, the state is still strongly entangled. This would make intuitive sense, given that the property responsible for the long-range correlations in quantum phase transitions is
entanglement.

 The plots of the results for the entanglement and the coherence correlator share a common structure, however the maximum values occur for different parameter values. This could mean that there is possibly another correlation that is more closely related to the entanglement between the atomic and molecular modes.

\section{Conclusion}

 We have argued here that in a system consisting of a  pair of tunnel-coupled BEC's, 
the individual bosons within the condensates cannot  be viewed as distinguishable subsystems. 
Subsequently entanglement in this system  should not be viewed as between
the individual bosons.  A more physically relevant description of this system is as a bipartite system where the subsystems are the two modes. Using this description we have analyzed quantitatively the entanglement between the two modes in the ground state of the coupled BEC's and its relation to the Mott insulator to superfluid phase transition. This idea was extended to consider the entanglement between the atomic and molecular modes in an atom-molecule BEC.

 On top of this, we have demonstrated that the dynamical scheme of \cite{mpe2}, argued to be viable with current state of the art technology, can be used to create a highly entangled state between the modes of the BEC system, over a smaller time-scale.

 As mentioned earlier, the amount of entanglement depends upon how the system is decomposed. In section \ref{dyn}, it was shown that the tunnel-coupled two-mode system can be viewed as a pseudo-angular momentum system - a single {\it qudit}. In this description - viewing the solely as a single qudit - it appears that there is no entanglement present in the system. Entanglement is only seen when the system is viewed in terms of subsystems, in this case, the two modes. In other words, the entanglement cannot be characterized when we neglect information about the underlying subsystems and only consider properties of the system as a whole.
 
 A possible way to create entanglement between individual bosons in the tunnel-coupled system would be to engineer some state within the condensate traps, then free the particles (see \cite{pu,LM} for examples of this procedure applied to other BEC systems). Once the bosons are free from the condensate they become
distinct allowing entanglement to form between them. However, whilst the bosons remain in condensate they are indistinguishable and cannot become entangled with
each other.\\

\acknowledgements

We would like to thank Tobias Osborne for insightful discussions. This work was supported by the Australian Research Council and an Australian Postgraduate Award to APH.


\begin{references}

\bibitem{mixse} C.H. Bennett, D.P. DiVincenzo, J.A. Smolin and W.K. Wootters, Phys. Rev. A {\bf 54} 3824 (1996).
 
\bibitem{man} M.A. Nielsen and I.L. Chuang, {\it Quantum computation and quantum information} (Cambridge University Press, Cambridge, 2000).

\bibitem{TON} T.J. Osborne and M.A. Nielsen, Phys. Rev. A {\bf 66}, 032110 (2002).

\bibitem{maphd} M.A. Nielsen, Ph.D. thesis, University of New Mexico (1998), quant-ph/0011036. 

\bibitem{cm1} X. Wang, H. Fu, and A.I. Solomon J. Phys. A {\bf 34}, 11307 (2001).
 
\bibitem{cm2} W.K. Wooters (2000), quant-ph/0001114.

\bibitem{cm3} M.C. Arnesen, S. Bose, and V. Vedral, Phys. Rev. Lett. {\bf 87}, 017901 (2001).

\bibitem{cm4} D.A. Meyer and N.R. Wallach (2001), quant-ph/0108104.

\bibitem{shi} Y. Shi (2002), cond-mat/0204058.

\bibitem{Benetal} C.H. Bennett, S. Popescu, D. Rohrlich, J.A. Smolin, and A.V. Thapliyal (2000), quant-ph/9908073.

\bibitem{zan} P. Zanardi, Phys. Rev. Lett. {\bf 87}, 077901 (2001).

\bibitem{leg} A.J. Leggett, Rev. Mod. Phys. {\bf 73}, 307 (2001).

\bibitem{GMJ} G.J. Milburn, J. Corney, E.M. Wright, and D.F. Walls, Phys. Rev. A  {\bf 55}, 4318 (1997).

\bibitem{bog} G.-S. Paraoanu, S. Kohker, F. Sols, and A.J. Leggett, J. Phys. B {\bf 34}, 4689 (2001).
 
\bibitem{natmae} A. S\o renson, L.-M. Duan, J.I. Cirac, and P. Zoller, {\it Nature} {\bf 409}, 63 (2001). 

\bibitem{mpe2} A. Micheli, D. Jaksch, J.I. Cirac, and P. Zoller (2002), cond-mat/0205369.

\bibitem{peres} A. Peres, {\it Quantum Theory: Concepts and Methods} (Kluwer Academic Publishers, Dordrecht, The Netherlands, 1993) p. 126.

\bibitem{li} Y.S. Li, B. Zeng, X.S. Liu, and G.L. Long, Phys. Rev. A, {\bf 64}, 054302 (2001).

\bibitem{VE} S.J. van Enk (2002), quant-ph/0206135.


\bibitem{atm1} P.D. Drummond, K.V. Kheruntsyan, and H. He, Phys. Rev. Lett. {\bf 81}, 3055 (1998).

\bibitem{atm2} K.V. Kheruntsyan and P.D. Drummond, Phys. Rev. A {\bf 61}, 063816 (2000).

\bibitem{atm3} J. Javanainen and M. Mackie, Phys. Rev. A {\bf 59}, R3186 (1999).

\bibitem{atm4} E. Timmermans, P. Tommansini, R. Cote, M. Hussein, and A. Kerman, Phys. Rev. Lett. {\bf 83}, 2691 (1999).

\bibitem{atm5} D.J. Heinzen, R.H. Wynar, P.D. Drummond, and K.V. Kheruntsyan, Phys. Rev. Lett. {\bf 84} 5029 (2000).

\bibitem{atm6} F.A. Abeelem and B.J. Verhaar, Phys. Rev. Lett. {\bf 83}, 1550 (1999).

\bibitem{zol} P. Zoller, {\it Nature} {\bf 417}, 493 (2002).

\bibitem{don} E.A. Donley et al., {\it Nature} {\bf 417}, 529 (2002).

\bibitem{fish} M.P.A. Fisher, P.B. Weichman, G. Grinstein, and D.S. Fisher, Phys. Rev. B {\bf 40} 546 (1989).

\bibitem{bh2} D. Jaksch, C. Bruder, J.I. Cirac, C.W. Gardiner, and P. Zoller, Phys. Rev. Lett. {\bf 81}, 3108 (1998).

\bibitem{vard} A. Vardi, V.A. Yurovsky, and J.R. Anglin, Phys. Rev. A {\bf 64}, 063611 (2001).

\bibitem{zhoa} H.-Q. Zhou, J. Links, R.H. McKenzie, and X.-W. Guan (2002), cond-mat/0203009.

\bibitem{zho} H.-Q Zhou, J. Links, and R.H. McKenzie (2002), cond-mat/0207540.

\bibitem{bell} J.S. Bell, Physics 1, 195 (1964).

\bibitem{dunn} J.A. Dunningham, S. Bose, L. Henderson, V. Vedra,l and K. Burnett, Phys. Rev. A {\bf 65}, 064302 (2002).

\bibitem{pres} J.Preskill, notes, \texttt{http://www.theory.caltech.edu/people}
\texttt{/preskill/ph229/\#lecture} (Chapter 5).

\bibitem{lee} H.W.Lee and J.Kim, Phys. Rev. A {\bf 63}, 012305 (2000).

\bibitem{fmot} M. Greiner, O. Mandel, T. Esslinger, T.W. H$\ddot{a}$nsch, and I. Bloch, {\it Nature} {\bf 415}, 39 (2002).

\bibitem{ost} A. Osterloh, L. Amico, G. Falci, and R. Fazio, {\it Nature} {\bf 416}, 608 (2002).

\bibitem{pu} H. Pu and P. Meystre, Phys.Rev.Lett {\bf 85}, 3987 (2000).

\bibitem{LM} L.-M. Duan, A. S\o renson, J.I. Cirac, and P. Zoller Phys.Rev.Lett {\bf 85}, 3991 (2000).

\end{references}
\end{document}